\documentclass[12pt]{article}

\usepackage{sbc-template}

\usepackage{graphicx,url}
\usepackage[T1]{fontenc}
\usepackage[utf8]{inputenc}  
\usepackage{xcolor}  
\usepackage{hyperref}  
\usepackage{algpseudocode}
\usepackage[Algoritmo]{algorithm}
\usepackage{bm}
\usepackage{listings} 
\usepackage{float}
\usepackage{subfigure}
\usepackage{enumerate}
\usepackage{amsmath}
\usepackage{setspace}
\usepackage{tabulary}
\newcolumntype{K}[1]{>{\centering\arraybackslash}p{#1}}

\usepackage{adjustbox}

\newcommand{\aspas}[1]{{``#1''}}
\mathchardef\mhyphen="2D

\newcommand{\mcode}[1]{{$\mathtt{#1}$}}

\title{Towards a Technique for Extracting Microservices\\from Monolithic Enterprise Systems}

\author{Alessandra Levcovitz\inst{1}, Ricardo Terra\inst{2}, Marco Tulio Valente\inst{1}}
\address{%
Universidade Federal de Minas Gerais (UFMG), Belo Horizonte, Brazil
\nextinstitute 
Universidade Federal de Lavras (UFLA), Lavras, Brazil
  \email{alessandralev@gmail.com,terra@dcc.ufla.br,mtov@dcc.ufmg.br}
} 

\begin{document} 

\maketitle

\begin{abstract} 
The idea behind microservices architecture is to develop a single large, complex, application as a suite of small, cohesive, independent services. On the other way, monolithic systems get larger over the time, deviating from the intended architecture, and becoming tough, risky, and expensive to evolve. This paper describes a technique to identify and define microservices on a monolithic enterprise system. 
As the major contribution, 
our evaluation demonstrate that our approach could identify good candidates to
become microservices on a 750 KLOC banking system, which reduced the size of the original system and took the benefits of microservices architecture, such as 
services being developed and deployed independently, and technology independence.
\end{abstract}

\section{Introduction}

Monolithic systems inevitably get larger over the time, deviating from their intended architecture and becoming hard, risky, and expensive to evolve~\cite{richardson14,sarkar09}. Despite these problems, enterprise systems often adopt monolithic architectural styles. Therefore, a major challenge nowadays on enterprise software development is to evolve monolithic system on tight business schedules, target budget, but keeping quality, availability, and reliability~\cite{markus00}.   

    Recently, microservices architecture emerged as an alternative to evolve monolithic applications~\cite{fowler14}.  The architecture proposes to develop a system as a set of cohesive services that evolve over time and can be independently developed and deployed. {Microservices} are organized as a  suite of small services where each one runs in its own process and communicates through lightweight mechanisms. These services are built around business capabilities and are independently deployed~\cite{newman15}. Thereupon, each service can be developed in the programming language that suits better the service characteristics, can use the most appropriate data persistence mechanism, can run on an appropriate hardware and software environment, and can be developed by different teams. There are many recent success stories on using microservices on well-known companies, such as Amazon and Netflix~\cite{microservices-amazon-netflix}.

	Therefore, microservices are an interesting approach to incrementally evolve enterprise application. More specifically, new business functions can be developed as microservices instead of creating new modules on the monolithic codebase. Moreover, existing components can be extracted from the monolithic system as microservices. This process may contribute to reduce the size of monolithic application, and create smaller and easier to maintain code.  

	In this paper, we 
	describe a technique to identify microservices on monolithic systems.  We successfully applied this technique on a 750~KLOC real-world monolithic banking system, which
	manages transactions from 3.5 million banking accounts and performs nearly 2 million authorizations per day. Our evaluation shows that the proposed technique is able to identify microservices  on monolithic system and that microservices can be a promising alternative to modernize legacy enterprise  monolithic system.  
	
	The remainder of this paper is organized as follows. Section~\ref{sec:technique} describes the proposed technique to identify microservices on monolithic systems. Section~\ref{sec:casestudy} evaluates the proposed technique on a real-world monolithic banking system. Section~\ref{sec:relatedwork} presents related work and Section~\ref{sec:conclusion} concludes the paper. 

\section{The Proposed Technique}
\label{sec:technique}

\begin{figure}[ht]
\centering
\includegraphics[width=11cm]{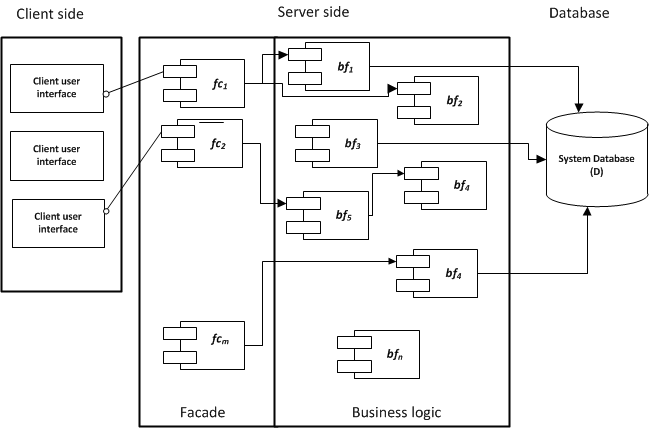}
\caption{Monolithic application}
\label{fig:monolithicapp}
\end{figure}

The proposed technique considers that monolithic enterprise applications have three main parts~\cite{richardson14}: a {\em client side} user interface, a {\em server side} application, and a {\em database}, as shown in Figure~\ref{fig:monolithicapp}. It also considers that a large system is structured on smaller subsystems and each subsystem has a well-defined set of business responsibilities ~\cite{conway68}. We also assume that each subsystem has a separate data store.

In formal terms, we assume that a system \mcode{S} is represented by a triple~\mcode{(F, B, D)}, where
\mcode{F = \{fc_1, fc_2, \dots, fc_{n'}\}} is a set of facades,
\mcode{B = \{bf_1, bf_2, \dots, bf_{n''}\}} is a set of business functions, and 
\mcode{D = \{tb_1, tb_2, \dots, tb_{n'''}\}}  is a set of database tables.
%
Facades~(\mcode{fc_i}) are the entry points of the system that call business functions~(\mcode{bf_i}). Business functions are 
methods that encode business rules and depend on database tables~(\mcode{tb_i}).
%
It is also important to define an 
enterprise organization \mcode{O = \{a_1, a_2, \dots, a_w\}} is divided into business areas \mcode{a_i}, each responsible for a business process. \\[-0.2cm]


\noindent 
We describe our technique to identify microservices on a system \mcode{S} in the following steps:\\

\begin{figure}[ht]
\centering
\includegraphics[width=8cm]{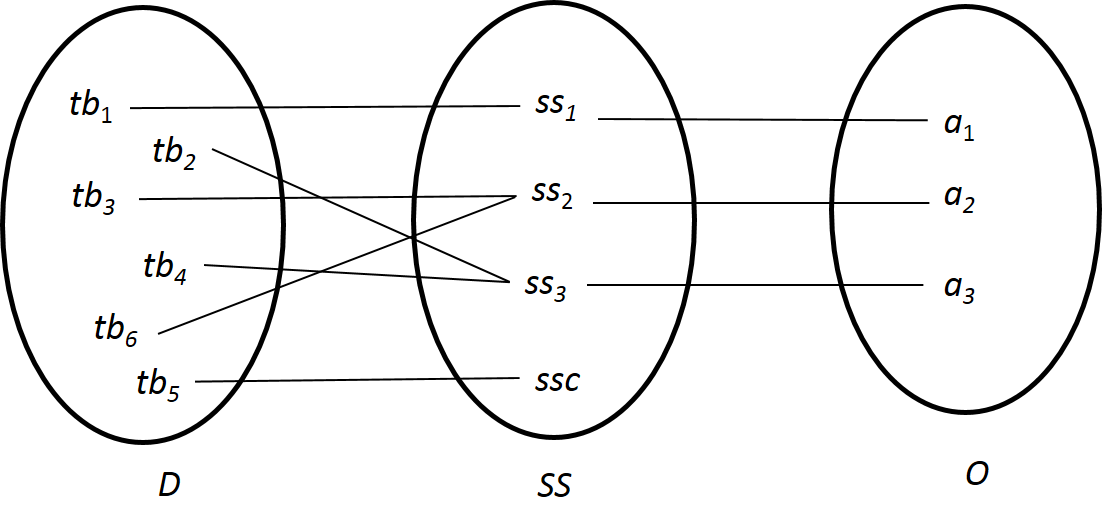}
\caption{Database decomposition}
\label{fig:database}
\end{figure}

\noindent {\bf Step \#1}: Map the database tables \mcode{tb_i \in D} into subsystems \mcode{ss_i \in SS}. Each subsystem represents a business area (\mcode{a_i}) of organization \mcode{O}.  
 For instance, as presented in Figure~\ref{fig:database}, subsystem~\mcode{SS_2}, which represents business area~\mcode{a_2}, depends on database tables~\mcode{tb_3} and \mcode{tb_6}. 
Tables unrelated to business process---e.g., error messages and log tables---are classified on a special subsystem called {\em Control Subsystem} (SSC).\\


\begin{figure}[ht]
\centering
\includegraphics[width=8cm]{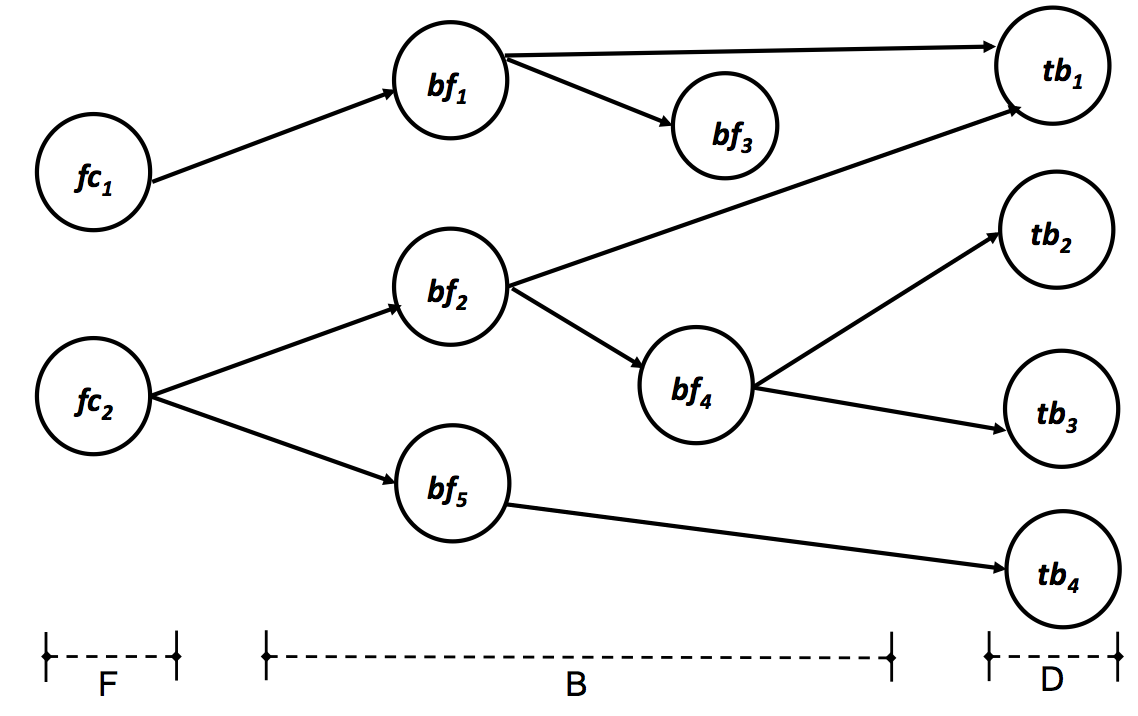}
\caption{Dependency Graph}
\label{fig:callgraph}
\end{figure}

\noindent {\bf Step~\#2}: Create a dependency graph $(V, E)$ where vertices represent facades (\mcode{fc_i \in F}), business functions (\mcode{bf_i \in B}), or database tables (\mcode{tb_i \in D}),
and edges represent: (i)~calls from facades to business functions; (ii)~calls between business functions; and (iii)~accesses from business functions to database tables. 
Figure~\ref{fig:callgraph} illustrates a graph where
facade~\mcode{fc_2} calls business function~\mcode{bf_2} (case~$i$),
business function~\mcode{bf_2} calls business function~\mcode{bf_4} (case~$ii$), and
business function~\mcode{bf_4} accesses database tables~\mcode{tb_2} and \mcode{tb_3} (case~$iii$).

\noindent {\bf Step \#3}:  Identify pairs \mcode{(fc_i, tb_j)}  where \mcode{fc_i \in F} and \mcode{tb_j \in D}, and there is a path from  \mcode{fc_i}  to \mcode{tb_j}  on the dependency graph. 
For instance, in the graph illustrated in Figure~\ref{fig:callgraph}, we identify the pairs \mcode{(fc_1, tb_1), (fc_2, tb_1), (fc_2, tb_2), (fc_2, tb_3), \textrm{and~} (fc_2, tb_4)}.\\

\noindent {\bf Step \#4}:  For each subsystem \mcode{ss_i} previously defined in Step~\#1, select pairs \mcode{(fc_i, tb_j)} identified on the prior step where \mcode{tb_j \in ss_i}. 
For instance, in our illustrative example, \mcode{ss_3 = \{tb_2, tb_4\}} then we select the pairs 
\mcode{(fc_2, tb_2)} and \mcode{(fc_2, tb_4)}.\\ 



\noindent {\bf Step \#5}:   Identify candidates to be transformed on  microservices. 
For each distinct pair \mcode{(fc_i, tb_j)} obtained on the prior step, we inspect the code of the facade and business functions that are on the path from vertex~\mcode{fc_i} to \mcode{tb_i} in the dependency graph. 
For instance, for pair~\mcode{(fc_2, tb_2)}, we inspect facade \mcode{fc_2} and business functions~\mcode{bf_2} and \mcode{bf_4}.
The inspection aims to identify which business rules actually depend on database table~\mcode{tb_j} and such operations should be described in textual form as rules.  
Therefore, for each pair \mcode{(fc_i, tb_j)}, a candidate microservice (\mcode{M}) is defined as follows: 

\begin{itemize}
	\item {\em Name}: the service name according to pattern \mcode{[subsystem name].[process name]}. For instance, \mcode{Order.TrackOrderStatus}.\\[-0.3cm]

	\item {\em Purpose}: one sentence that describes the main business purpose of the operation, which is directly associated to the accessed database entity domain. For instance, {\em track the status of a customer order}.\\[-0.3cm]



	\item {\em Input/Output}: the data the microservice requires as input and produces as output---when applied---or expected results for the operation. For instance, 
	microservice~\mcode{Order.TrackOrderStatus} requires as input the {\em Order Id} and
	produces as output a {\em List of finished steps with conclusion date/time} and a {\em List of pending steps with expected  conclusion date}. \\[-0.3cm]


	\item {\em Features}: the business rules the microservice implements, which are described as a verb (action), an object (related to the database table), and a complement. For instance, {\em Identify the order number}, {\em Get the steps for delivery for type of order}, {\em Obtain finished steps with date/time}, and {\em Estimate the date for pending steps}.\\[-0.3cm]


	\item	{\em Data}: the database tables the microservice relies on. \\[-0.3cm]
\end{itemize}

\noindent {\bf Step \#6}: Create API gateways to turn the  migration to  microservices transparent to clients.
%
API gateway consists of an intermediate layer between client side and server side application. It is a new component that handles requests from client side---in the same technology and interface as \mcode{fc_i}---and synchronizes calls to  the new microservice \mcode{M} and to \mcode{fc'_i}---a new version  of \mcode{fc_i}  without the code that was extracted and implemented on microservice  \mcode{M}. An API gateway should be defined for each facade.

		There are three cases of synchronization we have to consider in our evaluation: (i) when the input of \mcode{fc'_i} is the output of \mcode{M}  or the input of \mcode{M} is the output of \mcode{fc'_i};
	(ii)~when the input of \mcode{M} and \mcode{fc'_i} are the same as API gateway input and the instantiation order is irrelevant; and
	(iii)~when we have to split \mcode{fc_i} into two functions \mcode{fc'_i} and \mcode{fc''_i} and microservice \mcode{M} must be called after \mcode{fc'_i} and before \mcode{fc''_i}.
	
	%
	%
	
	
	On one hand, if we can synchronize the calls as described on case (i) or (ii), we identify the proposed microservice \mcode{M} as a \aspas{\em strong candidate}.
	On the other hand, if we can only synchronize the calls as in case (iii), we identify the proposed microservice as a \aspas{\em candidate with additional effort}.  
	Particularly in our technique, assuming a microservice of a subsystem~\mcode{ss_x}, 	if we identify a business rule in the microservice definition that needs to update data in \mcode{tb_i \in ss_x} and \mcode{tb_j \not\in ss_x} in the same transaction scope, we identify the proposed microservice as a \aspas{\em non candidate}.
	

	When every evaluated pair \mcode{(fc_i, tb_j)} of a subsystem is classified as microservice candidate, we recommend to migrate the entire subsystem to the new architecture. In this case, we have to implement the identified microservices, create an independent database with subsystem tables, develop API gateways, and eliminate the subsystem implementation (source code and tables) from system \mcode{S}. Although API gateways must be deployed in the same server as system \mcode{S} to avoid impacts on client side layer, the microservices can be developed using any technology and deployed wherever is more suitable. 

\section{Evaluation}
\label{sec:casestudy}
We applied our proposed technique on a large system from a Brazilian bank. The system handles transactions performed by clients on multiple banking channels (Internet Banking, Call Center, ATMs, POS, etc.). It has 750 KLOC in the C language and runs on Linux multicore servers. The system relies on a 
DBMS with 198 tables that performs, on average, 2 million transactions a day. 


\noindent {\bf Step \#1}: We identified 24 subsystems including subsystem~SSC.  Table~\ref{tb:ssandtables} shows a fragment of the result obtained after this initial step. 
Headers represents subsystems and their content represent the tables they rely on.
One problem we identified is that certain tables---highlighted in grey---%
are associated to more than one subsystem.

\begin{table}[ht]
\caption{Mapping of Subsystems and Tables}
\begin{center}
\begin{scriptsize}
\begin{tabular}{K{1cm}|K{1cm}|K{0.8cm}|K{0.8cm}|K{1.2cm}|K{1.2cm}|K{1.2cm}|K{1cm}|K{1.2cm}|K{1.1cm}}
\hline
{\bf Business Actions} & {\bf Service Charges} & {\bf Checks} & {\bf Clients} & {\bf Current Accounts}
& {\bf Saving Accounts} & {\bf Social Bennefits} & {\bf Pre-approved credit} & 
{\bf Debit and  Credit cards} & {\bf SMS Channel}\\\hline
ACO & AGT & CHS & CLT & \textcolor{gray}{\em CNT} & \textcolor{gray}{\em CNT} & BEN & LPA & CMG & CTS\\
ACB & ISE & TCE &  & CCT & CPO & DPB &  & INP & STS\\
RCA & PTC & ECH &  & \textcolor{gray}{\em RCC} & \textcolor{gray}{\em LAN} & DBC &  & NPP & CMS\\
 & PTF & HET &  & \textcolor{gray}{\em LAN} & \textcolor{gray}{\em RCC} & IBS &  & BIN & RCS\\
 & RTE & CCF &  & CCO & MPO & LBE &  & CCM & RLS\\
 & RTT &  &  & CCE & PPO &  &  & PBE & RTS\\
 & TPT &  &  & CHE & SPA &  &  &  & \\
 & TTE &  &  &  &  &  &  &  & \\
 & UTM &  &  &  &  &  &  &  & \\
 & DUT &  &  &  &  &  &  &  & \\
\hline
\end{tabular}
\end{scriptsize}
\end{center}
\label{tb:ssandtables}
\end{table}%

\begin{figure}[ht]
\centering
\includegraphics[width=8cm]{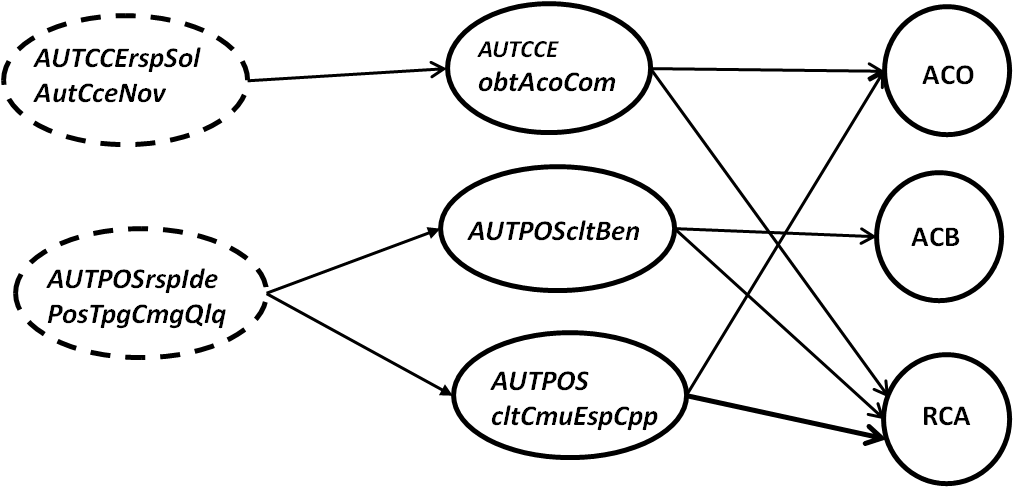}
\caption{Business Actions Subsystem Graph}
\label{fig:casestudygraph}
\end{figure}

\noindent {\bf Step \#2}: We created a dependency graph 
composed by 1,942 vertices (613 facades, 1,131 business functions, and 198 database tables), 5,178 edges representing function calls and 2,030 edges corresponding to database table accesses. Due to the size of our evaluated system, we proceed our evaluation to the following five subsystems: {\em Business Actions}, {\em Service Charges}, {\em Checks}, {\em Clients} and {\em SMS channel}. Table~\ref{tb:features} illustrates the characteristics of each subsystem. For subsystem {\em Business Actions}, we obtained the graph presented on Figure~\ref{fig:casestudygraph}.

\begin{table}[ht]
\caption{Evaluated Subsystems}
\begin{center}
\begin{small}
\begin{tabular}{l|K{1.5cm}|K{1.5cm}|K{1.5cm}|K{1.5cm}|K{1.5cm}}
\hline
{\bf Subsystem} & {\bf Business actions} & {\bf Service Charges} & {\bf Checks} & {\bf SMS channel} & {\bf Client}\\\hline
Tables (vertices) 	&3	&10	&5 &	6 &	1\\
Functions (vertices)	&5	&62&	29&	138&	>150\\
Function calls (edges)&	3&	79&	14&	133&	>150\\
Database accesses (edges)&	6&	14&	22&	140&	26\\
Microservices candidates&	1&	3&	8&	4&	4\\
\hline
\end{tabular}
\end{small}
\end{center}
\label{tb:features}
\end{table}%

\noindent {\bf Steps \#3--\#4}: Considering subsystem {\em Business Actions}, we find  the following pairs:
\begin{align*}
    \mathtt{(AUTCCErspSolAutCceNov, ACO)}&,&\mathtt{(AUTCCErspSolAutCceNov, RCA)} &, \\
    \mathtt{(AUTPOSrspIdePosTpgCmgQlq, ACO)}&,&\mathtt{(AUTPOSrspIdePosTpgCmgQlq, RCA)} &, \\
    \mathtt{(AUTPOSrspIdePosTpgCmgQlq, ACB)}&.
\end{align*}

\noindent {\bf Step \#5}: For pairs \mcode{(AUTCCErspSolAutCceNov, ACO), (AUTCCErspSolAutCceNov, RCA)} obtained on prior step, we inspect the code of facade \mcode{AUTCCErspSolAutCceNov} and the business functions it calls (\mcode{AUTCCEobtAcoCom}), and we identified the microservice described as follows:


\begin{itemize}
\item {\em Name}: \mcode{BusinessActions.ListBusinessActionsForAccount}.\\[-0.3cm]
\item {\em Purpose}: List business actions related to an account on the relationship channel.\\[-0.3cm]
\item {\em Input/Output}: Account number and channel id as input, and a list of business actions as output.\\[-0.3cm]
\item {\em Features}: Retrieve business actions assigned to the account; retrieve business actions enabled for the relationship channel; and retrieve the list of business actions assigned to the account and enabled for the channel.\\[-0.3cm]
\item {\em Data:} Database tables~ACO and RCA.
\end{itemize}

\noindent We also evaluated the other three pairs 
\mcode{(AUTPOSrspIdePosTpgCmgQlq, ACO)}, \mcode{(AUTPOSrspIdePosTpgCmgQlq, RCA),} \mcode{(AUTPOSrspIdePosTpgCmgQlq, ACB)}
and the business functions they call \mcode{AUTPOScltBen} and \mcode{ 
AUTPOScltCmuEspCpp}, and we identified the same microservices as the one described above. In fact, table~\mcode{ACB} could be merged with \mcode{ACO}.


\noindent {\bf Step \#6}: For facades \mcode{AUTCCErspSolAutCceNov} and \mcode{AUTPOSrspIdePosTpgCmgQlq}, we identified an API gateway that suits case ($i$) described in the proposed technique. 


\noindent We also evaluated steps \#3 to \#6 for subsystems: {\em Service Charges}, {\em Checks}, {\em Clients} and {\em SMS channel}. Although  subsystem {\em Service Charge} has 10 tables and 51 facades, we only identified and defined the following three microservices: \mcode{ServiceCharge.CalculateServiceCharge}, \mcode{ServiceCharge. IncrementServiceChargeUsage}, and \mcode{ServiceCharge.DecrementServiceChargeUsage}.
On one hand, we identified 14 API gateways that also suit case ($i$). On the other hand, we identified  other 37 API that suit case ($ii$). 
However, we can avoid the development of the last 37 API gateways since the called microservices have only input data and can be implemented with an asynchronous request.  Thus, particularly in this case, we suggest to use a message queue manager (MQM) for communication and substitute the C code that performs an update on database table for a \aspas{put} operation on a queue.


	For subsystems {\em Checks} and {\em SMS channel}, we identified microservices and APIs with the same characteristics of subsystem {\em Service Charge}, which indicates that both subsystems are good candidates to be migrated to microservices.  Nevertheless, for subsystem {\em Client}---which accesses only one table---we identified one microservice that must be called by more than 50 API gateways that suits case ($iii$).
Therefore, we did not recommend its migration to microservices, since  
the effort to split and remodularize more than 50 functions, create and maintain more than 50 API gateway are probably greater than the benefits of microservice implementation.  
		
Last but not least, we disregard subsystems that have one or more tables that appear in more than one subsystem list, such as table \mcode{CNT} of subsystem~{\em Current Account} (refer to Table~\ref{tb:ssandtables}) because our technique to identify microservices considers that only one subsystem handles operation on each table.  \\


\noindent {\bf Brief discussion:} We classify our study as well-succeeded because we could identify and classify all subsystems, and create and analyze the dependency graph that helped considerably to identify microservices candidates. As our practical result, we recommended to
migrate 4 out of the 5 evaluated subsystems to a microservice architecture.

\section{Related Work}
\label{sec:relatedwork}
Sarkar et al. described a modularization approach adopted on a monolithic banking system but they did not use a microservice architectural approach~\cite{sarkar09}.
Richardson evaluated a microservice architecture as a solution for decomposing monolithic applications. His approach considered the decomposition of  a monolithic system into subsystems by using  use cases or user interface actions ~\cite{richardson14}. Although this is an interesting  approach, in some situations a use case represents a set of operations of different business subsystems which are  synchronized by an actor action. Therefore, we do not always have an entirely system described with use cases. By contrast,
our technique starts by evaluating and classifying the database tables
into business subsystems, which demands access only to the source code
and the database model. Namiot et al. presented an overview of microservices architecture but they do not evaluate the concepts on a real-world system~\cite{namiot14}.
Terra et al. proposed a technique for extracting modules from
monolithic software architectures~\cite{terra8}. This technique is based
on a series of refactorings and aims to modularize concerns through
the isolation  of their code fragments. Therefore, they do not target
the extraction of modules that can be deployed
independently of each other, which is a distinguish feature of
microservice-based architectures.

\section{Conclusion and Future Work}
\label{sec:conclusion}

This paper describes a technique to identify microservices on monolithic systems. We successfully applied the proposed technique on a 750~KLOC real-world monolithic banking system, which demonstrated the
feasibility of our technique to identify upper-class microservices candidates on a monolithic system. The graph obtained for each subsystem helped considerably to evaluate the functions, identify, and describe microservices. 


There are subsystems that were not classified as good candidates to migrate to microservices, however. We found scenarios that would require a considerable additional effort to migrate the subsystem to a set of microservices. For instance, (i)~subsystems that share same database table, (ii)~microservice that represents an operation that is always in the middle of another operation, 
and (iii) business operations that involve more than one  business subsystem  on a transaction scope (e.g., money transfer from a check account to a saving account).

	
More important, the migration to the microservice architecture can be done 
incrementally. In other words, we can profit from the microservices architecture---e.g., services being developed and deployed independently, and technology independence--- without migrating the entire system to microservices.  Both kinds of systems architecture---monolithic and microservices---can coexist in a system solution. In fact, one challenge in using microservices is deciding when it makes sense to use it, which is exactly the ultimate goal of the technique proposed in this paper.


				As future work, we plan to evaluate other subsystems of the banking system, implement the identified  microservices, and measure in the field the real benefits. We also plan to  evaluate the proposed technique on Java monolithic systems.

\section*{Acknowledgment}
Our research has been supported by CAPES, FA\-PE\-MIG, and CNPq.

\bibliographystyle{plain}
\bibliography{terracopy}

\end{document}